\documentclass[aps,superscriptaddress,nofootinbib,twocolumn,prl]{revtex4-2}

\usepackage[T1]{fontenc}
\usepackage{soul}
\usepackage[usenames,dvipsnames]{color}

\usepackage{amsmath}
\usepackage{amssymb}

\usepackage{graphicx}

\usepackage{bm}
\usepackage{bbold}
\usepackage{mathrsfs}
\usepackage{slashed}

\newenvironment{frcseries}{\fontfamily{pzc}\selectfont}{}

\usepackage{dcolumn}
\usepackage{multirow}
\usepackage{hhline}
\usepackage{booktabs}
\usepackage{url}
\usepackage[colorlinks=true,
linkcolor=blue,anchorcolor=blue,citecolor=blue,filecolor=blue,menucolor=blue,runcolor=blue,
urlcolor=VioletRed
]{hyperref}

\usepackage[normalem]{ulem}
\usepackage{comment}

\newcommand{\derr}[2]{\frac{d{#1}}{d{#2}}}

\newcommand{\inparr}[2]{\partial{#1}/\partial{#2}}

\newcommand{\snn} {\sqrt{s_{_{\rm NN}}}}

\newcommand{\Tfo} {T_{\rm{fo}}}
\newcommand{\mufo} {\mu_{\rm{fo}}}
\newcommand{\Vfo} {V_{\rm{fo}}}
\newcommand{\mub} {\mu_B}
\newcommand{\mubc} {\mu_{B,{\rm{c}}}}
\newcommand{\Tc} {T_{\rm{c}}}
\newcommand{\gev} {~{\rm GeV}}
\newcommand{\mev} {~{\rm MeV}}
\newcommand{\fm} {~{\rm fm}}

\makeatletter
\renewcommand{\@fnsymbol}[1]{%
  \ifcase#1\or
    *\or
    **\or
    ***\or
    \dagger\or
    \ddagger
  \else
    \@ctrerr
  \fi
}
\makeatother

\usepackage{xcolor}

\begin{document}
	\title{
%Locating the critical point for the QCD phase transition from finite-size scaling of proton cumulants in heavy-ion collisions
Locating the QCD critical point with finite-size scaling of proton cumulants
}

\author{Agnieszka Sorensen}
\email{agnieszka.sorensen@gmail.com}
\affiliation{Institute for Nuclear Theory, University of Washington, Seattle, WA 98195, USA}
\affiliation{Facility for Rare Isotope Beams, Michigan State University, East Lansing, MI 48824, USA}

\author{Paul Sorensen}
\email{paul.sorensen@doe.gov}
\affiliation{U.S. Department of Energy, NP/Germantown Building, Washington, DC 20585, USA}

\begin{abstract}
We perform a finite-size scaling analysis of net-proton number cumulants in Au+Au collisions at center-of-mass energies between $\snn = 2.4$ and $200 \gev$ to search for evidence of a critical point in the QCD phase diagram. 
We use the second-order susceptibility and Binder cumulant, whose scaling with the system size is studied by using their dependence on the rapidity bin width~$W$.
We find that for collisions at $\snn \geq 7.7 \gev$, the susceptibility depends only weakly on~$W$ but increases as a power law in the baryon chemical potential~$\mu_B$, $(\mub-\mubc)^{-\gamma}$ with $\mubc=700\pm130$~MeV and $\gamma=1.6\pm0.5$, consistent with several recent theory estimates for the location of the QCD critical point.
We also find that the Binder cumulant appears to show a crossing within a similar range of~$\mu_B$ values. 
A comparable scaling behavior, however, is also found in dynamical simulations without critical fluctuations.
Moreover, we find that a Skellam distribution leads to apparent scaling of the second-order susceptibility when evaluated along the freeze-out line.
Simulations also show that the crossing of the Binder cumulant may be sensitive to changes in experimental acceptance at different beam energies. 
We conclude that while the observed scaling and crossing are consistent with the presence of a critical point, they cannot yet be taken as definitive evidence.
\end{abstract}

\maketitle

\section{Introduction}
\label{sec:introduction}

The phases and properties of strongly-interacting matter are important for understanding the early universe~\cite{Polyakov:1978vu,Susskind:1979up,Witten:1984rs}, nuclear structure~\cite{Brueckner:1955zze,Bogner:2009bt}, and interiors of neutron stars~\cite{Baym:1971ax,Collins:1974ky,Ozel:2016oaf}. 
Quantum chromodynamics~(QCD) is the established theory of this matter~\cite{Fritzsch:1973pi,Gross:1973id,Politzer:1973fx}, however, first-principles lattice QCD calculations at non-zero baryon chemical potential~$\mub$ are hindered by the fermion sign problem~\cite{Hasenfratz:1983ba,Cardy:1986gw}.
This precludes direct calculation of the QCD phase diagram at finite~$\mu_B$~\cite{Karsch:2001cy}.
While lattice QCD shows that the hadron to quark-gluon plasma~(QGP) transition at $\mub=0$ is a smooth crossover~\cite{Aoki:2006we,Ratti:2018ksb}, it is not known whether at finite~$\mub$ the nature of this transition changes and whether a critical point~(CP) exists where the crossover becomes a first-order phase transition~\cite{Asakawa:1989bq,Barducci:1989wi,Halasz:1998qr}. 
The search for this CP has been a high priority in nuclear physics for decades~\cite{Baym:2001in}.

Despite the sign problem, the CP location can be constrained using calculations at $\mub=0$~\cite{Allton:2002zi,Barbour:1997ej} or at imaginary $\mub$ through Lee-Yang edge singularities~\cite{Johnson:2022cqv,Rennecke:2022ohx}. 
The former exclude a CP for $\mub/T<3$~\cite{Bollweg:2022rps,Bollweg:2022fqq,Borsanyi:2020fev,Borsanyi:2022qlh}, where $T$ is the temperature, while the latter have recently yielded several CP predictions near $\mub\approx560$ MeV and $T\approx100$ MeV~\cite{Mukherjee:2019eou,Connelly:2020pno,Basar:2021hdf,Dimopoulos:2021vrk,Zambello:2023ptp,Schmidt:2022ogw}. 
Functional renormalization group and holographic gauge/gravity calculations also favor similar CP locations~\cite{Fu:2019hdw,Hippert:2023bel}.

The QCD phase diagram can be studied experimentally in heavy-ion collisions, where varying the collision energy~$\snn$ changes~$T$ and~$\mub$~\cite{Collins:1974ky,Asakawa:1989bq,Braun-Munzinger:1996nsn,Fukushima:2013rx,Sorensen:2023zkk}. 
The region containing recent theoretical CP predictions was probed by the STAR Collaboration's Beam Energy Scan~(BES) and Fixed-Target~(FXT) programs at the Relativistic Heavy Ion Collider~(RHIC)~\cite{STAR:2010vob,Odyniec:2019kfh,Meehan:2016qon}. 
Among proposed CP observables, significant attention is given to proton number cumulants and their beam-energy dependence~\cite{Stephanov:1998dy,Stephanov:1999zu,Asakawa:2009aj,Stephanov:2011pb}.
Here, we investigate the application of finite size scaling~(FSS) to search for the CP using %STAR~\cite{STAR:2021iop,STAR:2022etb} and HADES~\cite{HADES:2020wpc} 
experimental data. 
Although we find evidence consistent with a CP, similar signatures also arise in models without a CP.

\section{Finite-size scaling}
\label{sec:finite-size_scaling}

FSS emerges from invariance under a scale transformation~\cite{Ferdinand:1969zz}. 
Consider a system whose correlation length~$\xi$ diverges as $\xi\sim t^{-\nu}$, where $t = |(T-\Tc)/\Tc|$ is the reduced distance to the CP and $\nu$ the critical exponent.
If other length scales can be neglected, systems of different sizes~$L$ can be made self-similar by tuning~$t$ to reach matching values of reduced scale~$x = \xi/L \sim t^{-\nu}/L$.
This self-similarity occurs because, at equal~$x$, $\xi$ constitutes the same fraction of the system size~$L$, so that, up to an overall scale, systems at different~$L$ \textit{look the same}.
Next, if the system's susceptibility~$\chi$ diverges as $\chi \sim t^{-\gamma}$, where $\gamma$ is the associated critical exponent, one can express this dependence using~$x$ as $\chi \sim (xL)^{\gamma/\nu}$.
Consequently, plotting $\chi L^{- \gamma/\nu} \sim x^{\gamma/\nu}$ against~$x$ causes data corresponding to different~$t$ and~$L$ to fall onto a single \textit{universal curve}.
Crucially, this collapse occurs only for the correct~$x$, \textit{i.e.}, when the correct value of $T_c$ is used in~$t$.
FSS leverages this property to determine~$T_c$ empirically by varying its assumed value until data at different~$T$ and~$L$ collapse onto one curve.

Instead of~$x$, one often uses the scaling variable~$u \equiv t L^{1/\nu}$ for which $\chi \sim u^{-\gamma} L^{\gamma/\nu}$.
For a two-dimensional phase diagram the above considerations then give%\cite{??}
\begin{align}
\chi(L,t,h) = L^{\gamma/\nu}\Phi(tL^{1/\nu},hL^{\beta\delta/\nu}) ~,
\label{eq:chi_scaling}
\end{align}
where $t$ and $h$ are scaling fields measuring the distance from the CP, $\Phi$ is a universal scaling function, and $\gamma$, $\nu$, $\beta$, and $\delta$ are critical exponents. Equation~\eqref{eq:chi_scaling} implies that plotting $\chi L^{-\gamma/\nu}$ against $t L^{1/\nu}$ and $h L^{\nu/\beta\delta}$ should collapse data at different $L$, $t$, and $h$ onto a universal surface~$\Phi(t,h)$ (or a curve if either $t$ or $h$ is fixed).
As before, this collapse occurs only if correct values of~$\Tc$ and $\mubc$ enter~$t$ and~$h$. 
Thus, scanning~$\Tc$ and $\mubc$ for data collapse provides a method to locate the CP.

Scale invariance also implies another CP signature: the crossing of the Binder cumulant~\cite{binder1981}, 
\begin{align}
U_4 \equiv 1 - \frac{\langle (\delta N)^4\rangle}{3\langle (\delta N)^2\rangle^2}~.
\label{eq:Binder_cumulants}
\end{align}
$U_4$ was introduced in Ising model studies, where in the thermodynamic limit the order-parameter distribution is Gaussian in the high-temperature phase, giving $U_4=0$, and bimodal in the phase-coexistence region, giving $U_4 \to 2/3$. 
These limits are smoothed by finite-size effects, and near the CP $U_4$ scales as~$U_4(t,L) = f_U(tL^{1/\nu})$. 
Because $U_4$ carries no overall power of~$L$, at the CP $U_4(0,L)=f_U(0)$ is independent of system size.
Thus, curves of~$U_4(t)$ for different~$L$ cross near the CP, up to corrections to scaling. 
The crossing of~$U_4$ has been used in lattice QCD to identify critical behavior in the temperature and quark-mass plane~\cite{Fukugita:1990vu,Karsch:2001nf}.

In heavy-ion collisions, FSS has previously been applied to search for the QCD CP~\cite{Palhares:2009tf,Fraga:2011hi,Lacey:2016tsw}, with~$L$ varied by selecting different collision centrality intervals. 
These approaches have two main disadvantages: 
1) the connection of some scaled variables, such as HBT radii or mean-$p_T$ fluctuations, to~$\xi$, $\chi$, or the order parameter of the system is unclear, and 
2) varying centrality also changes the temperature, system shape, volume fluctuations, and dynamical evolution, potentially spoiling the analysis. 
Here, we instead analyze the net-proton susceptibility and vary~$L$ using subvolumes defined by the rapidity bin width, thus focusing on a well-motivated thermodynamic observable while avoiding changes in global system properties.

The applicability of FSS to heavy-ion experiments can be questioned based on
1)~the short collision lifetime, compounded by critical slowing
down~\cite{Berdnikov:1999ph,Nonaka:2004pg,Nahrgang:2018afz,Stephanov:2017ghc},
2)~non-critical correlations and fluctuations with their own dependence on the observation window~\cite{Skokov:2012ds,Steinheimer:2018rnd,Braun-Munzinger:2016yjz}, and 
3)~global baryon-number conservation, which also induces an acceptance dependence~\cite{Vovchenko:2020tsr,Braun-Munzinger:2020jbk,Kuznietsov:2022pcn}. 
Our analysis addresses these challenges by 1)~using system subvolumes, eliminating the need for the correlation length to reach the full system size~\cite{boxscaling} and reducing effects of baryon number conservation, and 
2)~using only central collisions, where non-critical correlations are suppressed because their contribution scales linearly with the system size while the number of particle pairs and typical backgrounds scale quadratically. 
Nevertheless, using rapidity bin width as a proxy for subvolume size is subject to a key caveat: rapidity bins do not always map cleanly onto isolated coordinate-space subvolumes. 
Thermal motion at freeze-out blurs the correspondence between coordinate- and momentum-space rapidity, introducing momentum-space correlations across roughly one unit of rapidity~\cite{Ling:2015yau,Ohnishi:2016bdf}, and particlization and hadronic decays can further alter the fluctuation signal and its acceptance dependence~\cite{Vovchenko:2020kwg,Bluhm:2016byc}.
%Nevertheless, using rapidity bin width as a proxy for subvolume size introduces another caveat, since thermal motion at freeze-out blurs the correspondence between coordinate- and momentum-space rapidity and can generate momentum-space correlations spanning about one unit of rapidity~\cite{Ling:2015yau,Ohnishi:2016bdf}. 
%Particlization and hadronic decays may further modify the fluctuation signal and its acceptance dependence~\cite{Vovchenko:2020kwg,Bluhm:2016byc}. 
%Therefore, rapidity bins may not correspond to cleanly separated coordinate-space subvolumes, potentially mixing their contributions within a given rapidity bin.

\section{Data}
\label{sec:experimental_data}

We extract the second-order susceptibility~$\chi_2$ and Binder cumulant~$U_4$ from 0--5\%-central Au+Au collisions at $\snn=2.4$--$200 \gev$. 
The $\snn = 7.7$--$200\gev$ data were measured by STAR in collider mode~\cite{STAR:2021iop}, while the $\snn =2.4$ and $3.0 \gev$ data come from fixed-target measurements by HADES~\cite{HADES:2020wpc} and STAR~\cite{STAR:2022etb}, respectively.
Varying $\snn$ changes the chemical freeze-out parameters~$(\Tfo,\mufo)$ and thus the distance from the CP. 
Here, we use $(T,\mu_B)$ and $(\Tfo,\mufo)$ interchangeably.

We vary the system size by selecting system subvolumes defined by different rapidity bin widths~$W$, \textit{i.e.}, rapidity \textit{windows}~\cite{boxscaling}, and use central collisions to minimize volume fluctuation corrections.
However, at $\snn = 2.4$ and $3.0 \gev$ the net-baryon density and particle ratios vary considerably with~$W$~\cite{STAR:2023uxk}, increasing systematic uncertainty at those energies. 
Moreover, at $\snn = 3.0 \gev$ only asymmetric~$W$ bins are available, with the largest window extending from $y=0$ to beam rapidity to match the range in~$W$ used at other energies.

Using the second-order net-proton cumulants~$C_2$ and chemical freeze-out parameters~$T_{\rm fo}$ and~$\Vfo$ at different~$\snn$ and~$W$, we extract $\chi_2$ as
\begin{align}
\chi_{2}(W,\mufo) = \frac{C_2(W, \mufo)}{\Tfo^{3} W d\Vfo / dy} ~.
\label{eq:susceptibility_extracted}
\end{align}
Here, $T_{\rm fo}$ and~$d\Vfo/dy$ are assumed independent of~$W$.
Technical details of this extraction, including challenges at low beam energies, are discussed in Appendix~\ref{sec:extracting_susceptibilities}.
We also extract the Binder cumulant as
\begin{align}
U_4 %= 1 - \frac{\langle (\delta N)^4\rangle}{3\langle (\delta N)^2\rangle^2}
= \frac{-C_4}{3C_2^2} ~.
\end{align}

\section{Results}
\label{sec:results}

\begin{figure}[t!] % t = top, b= bottom, h = here, ! = enforce
	\centering
	\includegraphics[width=0.49\textwidth]{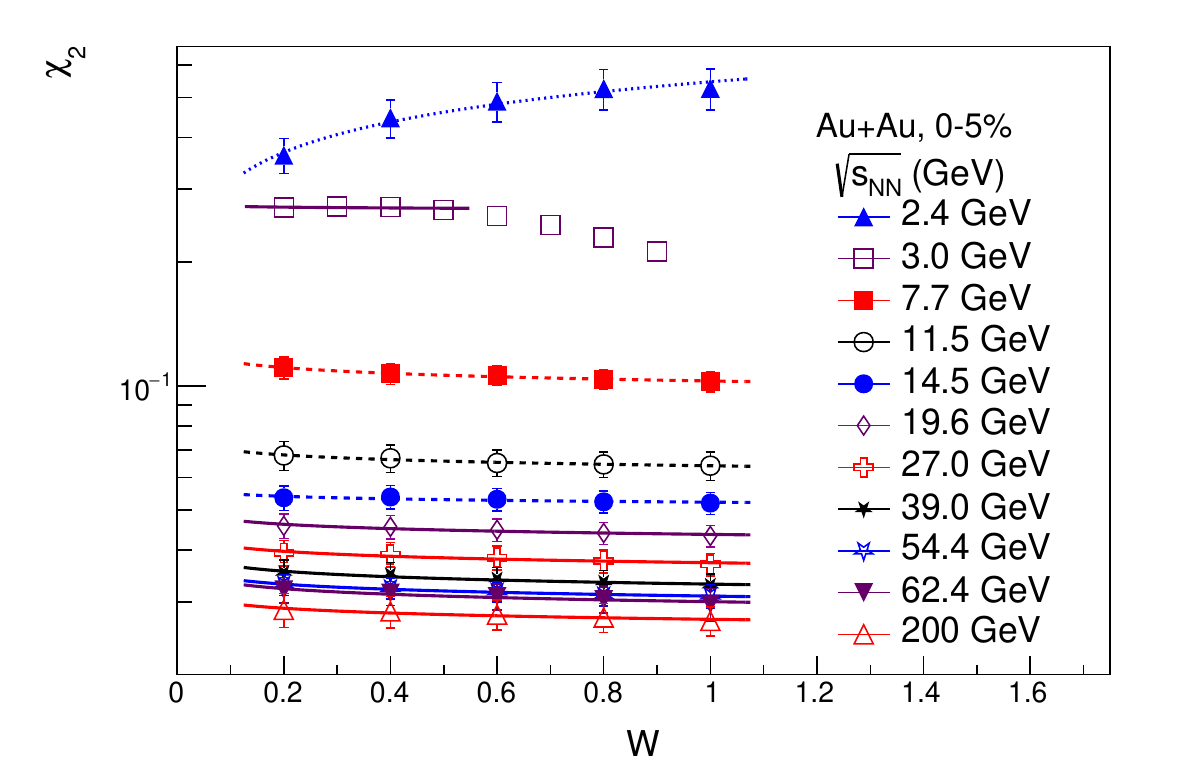} \\
	\includegraphics[width=0.49\textwidth]{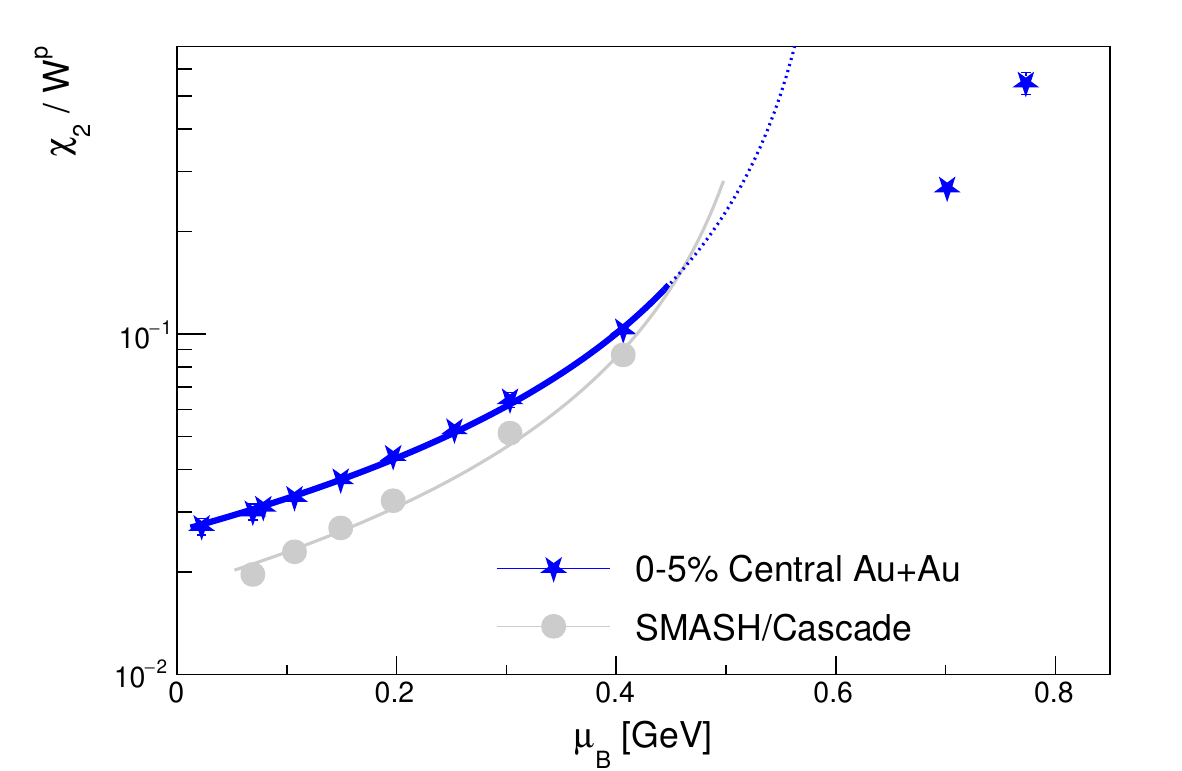} 
 	\caption{\textit{Top:} Susceptibility $\chi_2$ versus~$W$. 
    Curves show power-law fits to data with $\chi_2=aW^p$. 
    \textit{Bottom:} The $\mub$-dependence of $a=\chi_2/W^p$ with $p=-0.04$. Gray circles show results from \texttt{SMASH} simulations without a CP or mean fields.
	}
	\label{fig:one}
\end{figure}

The top panel of Fig.~\ref{fig:one} shows the extracted~$\chi_2$ as a function of~$W$ for $\snn=2.4$--$200 \gev$.  
Overall, $\chi_2$ decreases with beam energy. 
For $\snn = 7.7 \gev$ and above, $\chi_2$ depends only weakly on~$W$, decreasing slightly as~$W$ increases. 
This nearly flat behavior is consistent with either 
1)~short-range or weak rapidity correlations, or 
2)~thermal motion and hadronization mixing particles across rapidity windows~\cite{Ling:2015yau}. 
The weak decrease with~$W$ is also consistent with baryon number conservation effects~\cite{Vovchenko:2020tsr}. 
At $\snn = 2.4$ and $3.0 \gev$, the stronger~$W$-dependence likely reflects the absence of an extended central rapidity region, since at low~$\snn$ beam rapidity lies close to midrapidity. 
This complicates the use of~$W$ as a FSS size parameter for low~$\snn$, especially at $\snn = 3.0 \gev$, where only asymmetric windows ($0<y<W$) are available. 
For the remaining energies, the analysis uses symmetric ($|y|<W/2$) windows.

\textbf{Dependence of susceptibility~$\chi_2$ on $W$ and $\mu_B$.} 
To understand the behavior of~$\chi_2$, we first fit $\chi_2(W)$ with a power law, $\chi_2 = aW^p$, excluding the largest~$W$ values for $\snn = 3.0 \gev$ where the data extend to beam rapidity. 
Except at $\snn = 2.4 \gev$, we find that the~$W$-dependence of~$\chi_2$ is a power law with $p = -0.04$. 
The uncertainty on~$p$ is difficult to quantify because correlations among the systematic errors are unknown. 
Adding systematic and statistical errors in quadrature makes $p$ consistent with zero, while statistical errors alone give an uncertainty below 0.001. 
At $\snn = 2.4 \gev$, the data instead requires a positive exponent, $p=0.025$.

We next examine the $\mu_B$-dependence of the fit parameter~$a =\chi_2/W^p$, shown in the bottom panel of Fig.~\ref{fig:one}.
If $a$ also follows a power law, $a\sim t^{-\gamma}$, then $\chi_2 \propto t^{-\gamma}W^{p}$ is homogeneous in~$t$ and~$W$.
Any homogeneous function can be rescaled, collapsing data at different~$t$ and~$W$ onto one curve -- \textit{i.e.}, homogeneity of $\chi_2$ implies scaling.
We indeed find that at $\mub \lesssim 400$, $\chi_2/W^p$ is well-described by a power law in~$t$, diverging at~$\mubc=700\pm130 \mev$.
The fitted exponent~$\gamma=1.6\pm0.5$ is consistent with the 3-D Ising value $\gamma_{\rm 3DI}=1.237$~\cite{2009JSP}. 
Fixing $\gamma = \gamma_{\rm 3DI}$ gives~$\mubc=605\pm11 \mev$, while the mean-field value $\gamma_{\rm MF} =1$ yields~$\mubc=550\pm10 \mev$. 
Unknown point-to-point correlations of systematic uncertainties complicate drawing conclusions from the fits' $\chi^2$ values, however, in all cases $\chi^2$ per degree of freedom is well below 1. 
The $\snn = 2.4$ and $3.0 \gev$ data \mbox{do not follow this power law.}

If FSS applies to the data at $\mub \lesssim 400$ MeV, the observed power law implies a CP at $t \sim \mu_c - \mu_B  = 0$. 
However, this interpretation assumes that over the BES range $\chi_2$ is dominated by critical fluctuations with little or no background. 
To assess non-critical contributions, Fig.~\ref{fig:one} also shows the same analysis for \texttt{SMASH} hadronic transport simulations~\cite{Weil:2016zrk,Hammelmann:2022yso}, run without a CP or mean fields. 
Despite containing no critical fluctuations, \texttt{SMASH} produces a similar power-law increase in~$\chi_2$, diverging at~$\mub = 558 \mev$, calling into question whether the scaling of experimental data provides evidence for a~CP.

\begin{figure}[t!] % t = top, b= bottom, h = here, ! = enforce
	\centering
	\includegraphics[width=0.49\textwidth, trim={0.25cm 0.25cm 0.25cm 0.75cm}, clip]{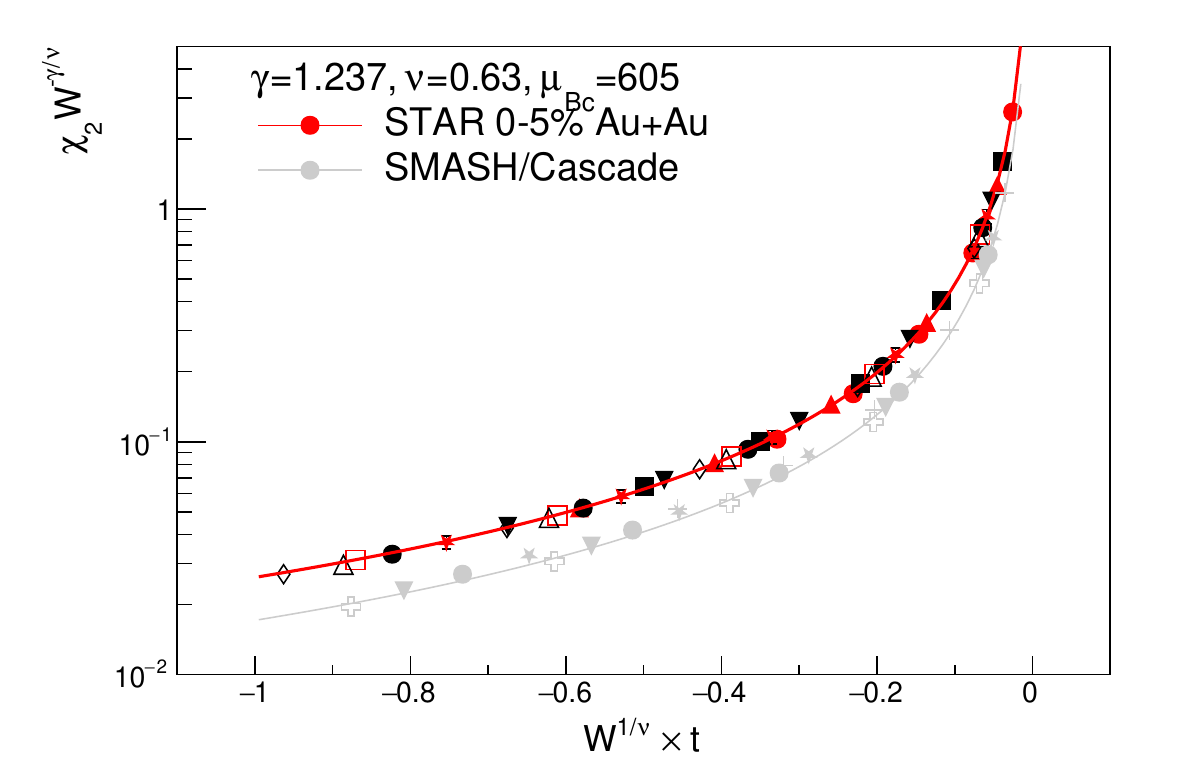}
 	\caption{A FSS scaling analysis of STAR $\chi_2$ data and a non-critical reference based on \texttt{SMASH}.
	}
	\label{fig:two}
\end{figure}

\textbf{FSS analysis of susceptibility.}
Figure~\ref{fig:two} shows the FSS scaling of $\chi_2W^{-\gamma/\nu}$ versus $tW^{1/\nu}$, where $t = \mu_B/\mu_c - 1 $ and we use the 3-D Ising critical exponents $\gamma_{\rm 3DI}$ and $\nu_{\rm 3DI}=0.63$~\cite{2009JSP}, obtaining $\mubc = 605 \mev$. 
As anticipated given the analysis above, the data exhibit very good scaling for $\snn \geq 7.7 \gev$, with the scaling function well-described by a power law, $\chi_2W^{-\gamma/\nu}=A(tW^{1/\nu})^b$ where $A=0.0261\pm0.0005$ and $b=-1.26\pm0.01$. 
Although the weak $W$-dependence of $\chi_2$ suggests modest finite-size effects, the power-law behavior of~$\Phi(tW^{1/\nu})$ is consistent with its expected tail far from the CP.
There, the correlation length is much smaller than the system size and $\chi_2=W^{\gamma/\nu}\Phi(tW^{1/\nu})$ should be independent of $W$, requiring that $\Phi(tW^{1/\nu})\sim (tW^{1/\nu})^{-\gamma}$. 
Thus, the data may reflect the tail of~$\Phi$, with the divergence of~$\chi_2(t)$ still pointing to the CP despite the weak $W$-dependence. 
However, Fig.~\ref{fig:two} also shows a scaling curve from \texttt{SMASH} in cascade mode. 
Although the scaling is not as good as for the STAR data, it demonstrates that a model without a CP or critical fluctuations also exhibits apparent FSS and power-law behavior over the same $\snn$ range.
Thus, while the observed scaling is compatible with a CP, the simulation results show that it may also be coincidental.

If the scaling is accidental, non-critical contributions to~$\Phi$ must \textit{mimic} critical behavior.
Using $u = t W^{1/\nu}$ together with Eqs.~\eqref{eq:chi_scaling} and~\eqref{eq:susceptibility_extracted}, we write $\Phi(u) = u^{-\gamma} \chi_2 t^{\gamma} = u^{-\gamma}  F(W, \mufo)$, where the ``residual factor'' is
\begin{align}
F(W, \mufo) %= \chi_2 W^{-\gamma/\nu} 
%= \frac{C_2(W, \mufo)}{\Tfo^{3} V(W)} W^{-\gamma/\nu}
= \frac{C_2(W, \mufo) n_P(W, \mufo)}{ C_1(W, \mufo) \Tfo^{3}} \Big(1 - \frac{\mufo}{\mu_c} \Big)^{\gamma}
\label{eq:non-critical_part_of_the_scaling_function}
\end{align}
and $n_P$ is the net-proton density.
Thus, $\Phi(u)$ appears as a power law when $F  \approx \rm{const}$ over the considered range of~$u$; as before, this depends on the assumed value of~$\mu_c$.

\begin{figure}[t!] % t = top, b= bottom, h = here, ! = enforce
	\centering
	\includegraphics[width=0.48\textwidth, trim={0.00cm 0.0cm 0.0cm 0.1cm}, clip]{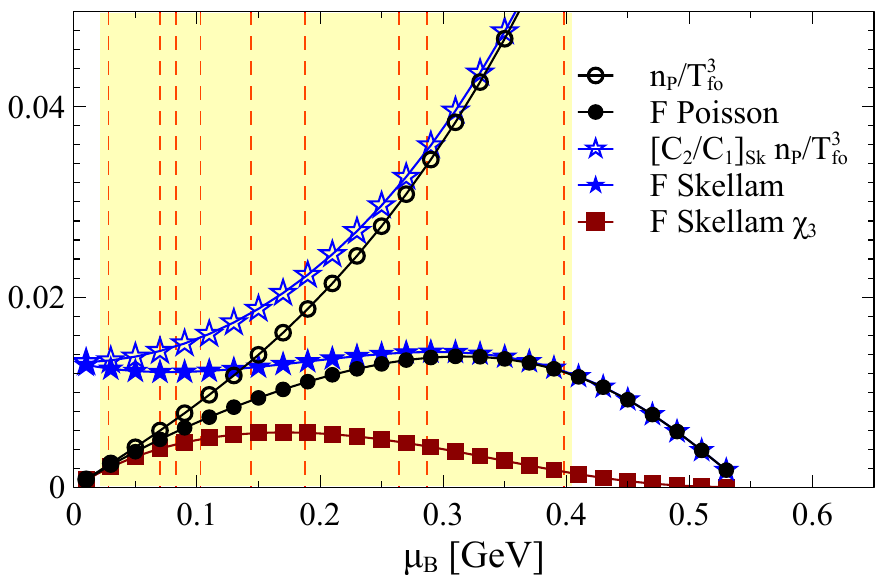} 
 	\caption{$F$ for $\chi_2$ from Poisson (solid black circles) and Skellam (solid blue stars) distributions, and for Skellam~$\chi_3$ (maroon squares), using $\mu_c = 0.55 \gev$.
    Open symbols show $(C_2/C_1)n_P/T_{\rm fo}^3$ for Poisson (equivalent to $n_P/T_{\rm fo}^3$) and Skellam~$\chi_2$%; Skellam~$\chi_3$ is identical to Poisson
    .
    Yellow shading marks the BES $\mub$ range, dashed red lines mark measured $\mufo$ values.
	}
	\label{fig:accidental_scaling}
\end{figure}

Eq.~\eqref{eq:non-critical_part_of_the_scaling_function} allows us to test whether $F \approx \rm{const}$ along the freeze-out line for simple non-critical contributions.
We consider Poisson and Skellam distributions, with details given in Appendix~\ref{sec:non-critical_scaling}.
We determine an optimal global~$\mu_c$ for each distribution by minimizing the relative standard deviation of~$F$ over the BES range, obtaining $\mu_c^{(\rm P)} = 0.48\gev $ and $\mu_c^{(\rm Sk)} = 0.55 \gev$ for Poisson and Skellam, respectively.
Since these values are close, we use $\mu_c = \mu_c^{(\rm Sk)}$ for each to simplify the comparison. 
Figure~\ref{fig:accidental_scaling} shows both~$\chi_2(\mufo) = (C_2/C_1)n_P/T_{\rm fo}^3$ and the resulting~$F(\mu_B)$.
The former rises sharply at large~$\mu_B$ for both Poisson and Skellam, which eventually become indistinguishable.
A striking difference between the two cases appears in~$F(\mu_B)$: $\sigma(F)/\langle F\rangle \approx 44\%$ for Poisson but only~$7\%$ for Skellam; using $\mu_c^{(\rm P)}$ lowers the Poisson value only to~$41\%$.
Thus, the Skellam~$F$ is nearly flat in~$\mu_B$ and can closely mimic the expected scaling over the full BES range.
This can be well understood: at large~$\mu_B$, 
the power-law behavior of $\chi_2(\mufo)$ is largely generated by the freeze-out line, while for $\mu_B \to 0$, the rise in $(C_2/C_1)_{\rm Sk}$ compensates for the decrease in $n_P/T_{\rm fo}^3$, creating a power-law behavior across the whole range. 
Since BES net-proton multiplicity distributions are well described by Skellam statistics~\cite{Braun-Munzinger:2014lba}, this may explain why apparent scaling occurs in experiment and in simulations that reproduce proton and antiproton multiplicities.

\begin{figure}[b!] % t = top, b= bottom, h = here, ! = enforce
	\centering
	\includegraphics[width=0.48\textwidth, trim={0.75cm 0.1cm 6.0cm 0.5cm}, clip]{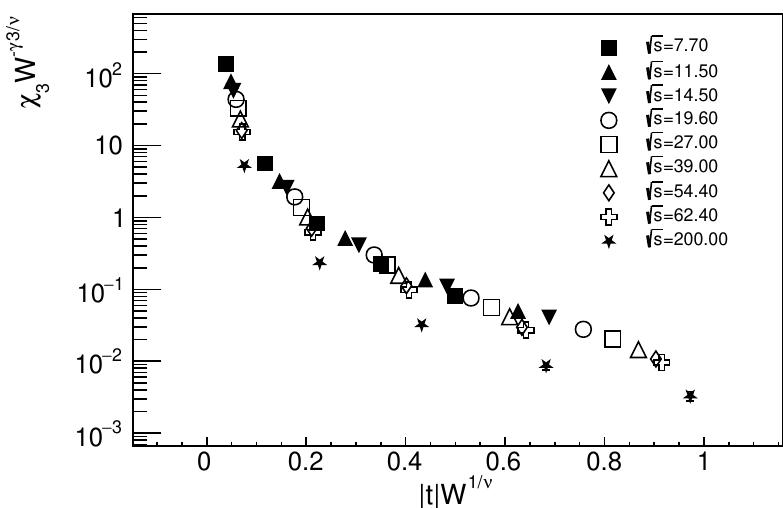}
 	\caption{FSS analysis of STAR~$\chi_3$ data, using $\gamma_{3, \rm{3DI}} = 2.8$, with the extracted $\mu_c = 811 \mev$. 
	}
	\label{fig:scaling_of_chi3}
\end{figure}

A similar analysis applies to higher-order susceptibilities.
For a Skellam distribution, even orders have the same functional form as~$\chi_2$, differing only in critical exponents, while odd orders follow Poisson.
The Skellam $\chi_3$ results are shown in Fig.~\ref{fig:accidental_scaling}. 
The larger critical exponent associated with~$\chi_3$ yields larger~$\mu_c^{\rm (Sk3)} = 0.65\gev$ (see Appendix~\ref{sec:non-critical_scaling}).
We find $\sigma(F)/\langle F \rangle \approx37\%$ for $\mu_c^{(\rm Sk)} = 0.55 \gev$, reduced only to $31\%$ at $\mu_c^{\rm (Sk3)}$, suggesting that scaling is less likely for~$\chi_3$. 
Indeed, FSS analysis of STAR~$\chi_3$ data, shown in Fig.~\ref{fig:scaling_of_chi3}, yields~$\mu_c = 811 \mev$, well above the~$\chi_2$ value. 
Moreover, the data does \textit{not} scale well.
We find similar results for \texttt{SMASH}.

Extending our analysis to both~$t$ and~$h$ leaves our conclusions largely unchanged.
The data show little sensitivity to~$T_c$, supporting our focus on scaling \mbox{with~$\mu_B$ alone.}

\begin{figure}[t!] % t = top, b= bottom, h = here, ! = enforce
	\centering
	\hspace{3.0mm}\includegraphics[width=0.495\textwidth, trim={0.0cm 0.0cm 0.55cm 0.0cm}, clip]{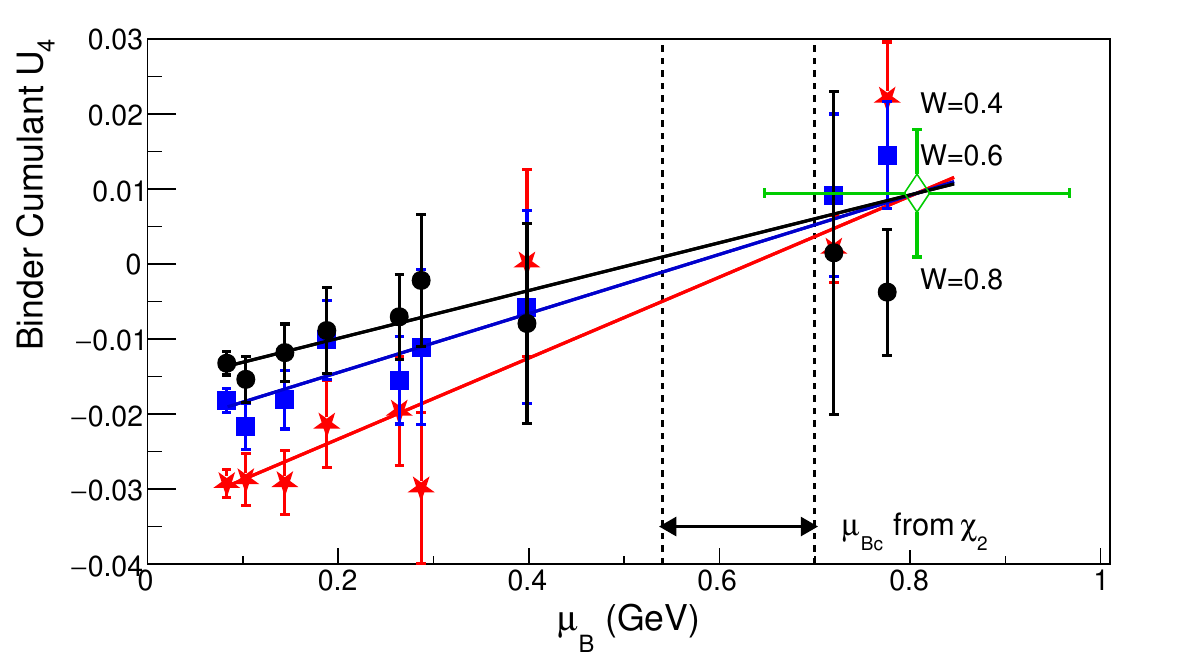} \\
    \includegraphics[width=0.485\textwidth]{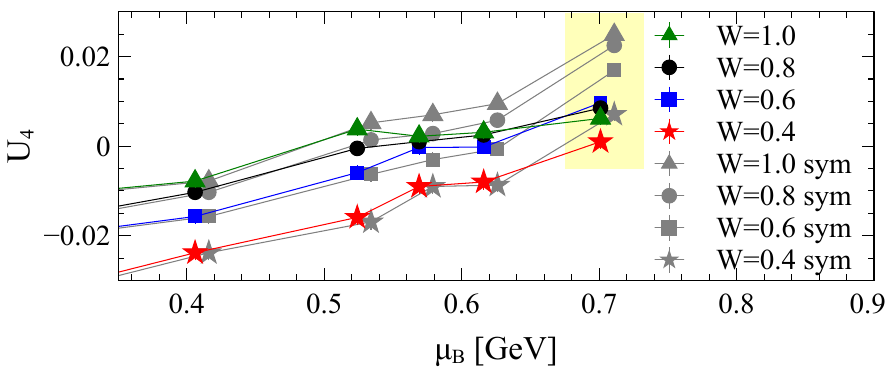} 
 	%\captionsetup{justification=raggedright}
 	\caption{$U_4$ versus $\mub$ for $W = 0.4$ (red stars), 0.6 (blue squares), and 0.8 (black dots).
    \textit{Top:} Lines show a simultaneous fit to all data, green open diamond marks the fitted crossing point.
    \textit{Bottom:} \texttt{SMASH} results for asymmetric (color marks) and symmetric (grey marks, $\mu_B$-shifted for clarity) $W$ for FXT data.
    Yellow rectangle highlights largest differences.
	}
	\label{fig:Binder_cumulant}
\end{figure}

\textbf{Binder cumulant.}
We also investigate the crossing of $U_4(\mub)$ at fixed~$W$, which uses higher-order cumulants and requires no thermal model input.
The top panel in Fig.~\ref{fig:Binder_cumulant} shows $U_4(\mub)$ for $W = 0.4$, 0.6, and 0.8. 
At low $\mub$, $U_4$ is consistent with a Skellam distribution. 
The $W=0.8$ data rise slowly with~$\mub$, while the $W = 0.4$ and 0.6 data rise faster, starting below the $W=0.8$ curve for $\mub<400$ and ending above it for $\mub>720$ MeV. 
A joint linear fit to the data, with three slopes and crossing coordinates~$(\mubc, U_0)$ as fit parameters, gives $\chi^2/{\rm d.o.f.} =  13/23$ and a crossing at $\mubc=807 \pm 16 \mev$ and $U_0=9.4\pm0.3 \times 10^{-3}$. 

This potential crossing relies on the $\snn = 2.4$ and $3.0 \gev$ data, where rapidity dependence is strong and the STAR $\snn = 3.0 \gev$ data uses asymmetric bins ($0<y<W$). 
We again compare this observation to a non-critical baseline using \texttt{SMASH}. Although no crossing appears in the simulations with symmetric bins, using asymmetric bins in the FXT region as done in experiment seems to produce an apparent crossing.
The bottom panel of Fig.~\ref{fig:Binder_cumulant} shows this most clearly at $\snn = 3.0 \gev$ where $U_4$ from asymmetric~$W$ bins cluster together, unlike those from symmetric bins.
An analysis with symmetric $W$ bins is needed before interpreting this signal.

Appendix~\ref{sec:details_of_codes_and_data} details codes and synthetic data used in this analysis.

\section{Summary}

In this work, we investigate a novel finite-size scaling analysis of net-proton number distributions from heavy-ion collisions, using the rapidity window~$W$ as the system-size scale.
We test the collapse of scaled susceptibilities and Binder cumulant crossing. 
Both methods indicate a CP near $\mu_B = 700 \pm 100 \mev$, consistent with recent theoretical predictions, however, neither provides definitive evidence for critical behavior.

The~$\chi_2$ data show a weak $W$-dependence, but exhibit a good scaling collapse and a power law in~$\mub-\mubc$ with an exponent consistent with the \mbox{3-D} Ising universality class. 
However, similar behavior arises in hadronic cascade simulations without a CP.
We show that this apparent scaling can emerge non-critically: along the BES chemical freeze-out line, cumulants of a Skellam distribution can mimic the expected critical power law.
Thus, the apparent FSS of~$\chi_2$ is not unique to critical behavior.
This effect is weaker for~$\chi_3$, for which data and simulations show poorer scaling and favor larger~$\mubc$.

The experimental data show an apparent crossing in the Binder cumulant, but simulations without a CP reproduce a similar pattern when using the asymmetric rapidity acceptance at $\snn=3.0 \gev$.
Thus, the crossing is sensitive to acceptance changes across beam energies and cannot presently provide strong evidence for a CP.

We note that the ``non-critical baselines'' employed here contain non-trivial QCD physics: the chemical freeze-out line reflects the system's thermodynamic evolution, while the hadronic spectrum entering thermal models and simulations follows directly from QCD. 
Moreover, while net-proton fluctuations are approximately Skellam-like, it does not by itself imply that the scaling is trivial, but calls for a better understanding of how the ingredients of the scaling function evolve across the phase diagram.

\section{Acknowledgements}

The authors would like to thank Vladimir Skokov, Larry McLerran, and Scott Pratt for their comments on the manuscript. 
The authors would also like to thank the participants of the ``The QCD Critical Point: Are We There Yet?'' program at the Institute for Nuclear Theory~(INT), University of Washington, for many insightful discussions, with special thanks to Larry McLerran, Gy\H{o}z\H{o} Kov\'acs, Mikhail Stephanov, and Volker Koch.
We thank the INT for its kind hospitality and stimulating research environment, supported by the U.S.\ Department of Energy grant No.\ DE-FG02-00ER41132.

P.S.\ would like to thank Eduardo Fraga and Leticia Palhares for fruitful collaborations on previous FSS analyses.
A.S.\ would like to thank Jacquelyn Noronha-Hostler for access to high-performance computing resources, and acknowledges use of the University of Illinois Urbana-Champagin Campus Cluster. 
A.S.\ also acknowledges that this research used resources of the National Energy Research Scientific Computing Center~(NERSC), a DOE Office of Science User Facility supported by the Office of Science of the U.S.\ Department of Energy under Contract No.\ DE-AC02-05CH11231 using NERSC award NP-ERCAP0037051.
A.S.\ further acknowledges support by the U.S.\ Department of Energy, Office of Science, Office of Nuclear Physics, under Grant No.\ DE-FG02-00ER41132 during the initial stages of this research.

\bibliography{cumulants_finite_size_scaling,noninspire}

\newpage

\appendix
\setcounter{secnumdepth}{2}

\section{Extraction of susceptibilities}
\label{sec:extracting_susceptibilities}

The cumulants, central moments, and susceptibilities are related by
\begin{align}
C_1 &= \big\langle N \big\rangle = VT^{3}\chi_1 ~, 
\label{eq:cumulants_1} \\
C_2 &= \big\langle (\delta N)^2 \big\rangle = VT^3\chi_2 ~,
\label{eq:cumulants_2} \\
C_3 &= \big\langle (\delta N)^3 \big\rangle = VT^3\chi_3 ~,
\label{eq:cumulants_3} \\
C_4 &= \big\langle (\delta N)^4 \big\rangle - 3\big\langle (\delta N)^2\big\rangle ^2 = VT^3\chi_4 ~.
\label{eq:cumulants_4}
\end{align}
Here, $C_i$, $\langle (\delta N)^i\rangle $, and $\chi_i$ denote the $i$th order cumulant, central moment, and susceptibility, respectively, and $V$ is the volume.
Cumulant measurements are often reported as ratios, for which the common factor $VT^3$ cancels, leaving ratios of susceptibilities. 
To extract $\chi_2$, we obtain $VT^3$ from thermal-model parametrizations of the chemical freeze-out temperature~$\Tfo$ and volume per unit rapidity~$d\Vfo/dy$.
Equation~\eqref{eq:susceptibility_extracted} then follows with $V = W d\Vfo / dy$; analogous expressions apply to higher-orders.
For $\snn \geq 7.7 \gev$, we use net-proton distributions, while at $\snn=2.4$ and $3.0 \gev$ antiproton production is negligible and proton distributions suffice.

While Ref.~\cite{Andronic:2017pug} parametrizes~$\Tfo$ and~$\mufo$, $d\Vfo/dy$ is not typically parameterized.
Chemical freeze-out volumes per unit rapidity have, however, been extracted from statistical-model fits to hadron yields over a broad energy range~\cite{Andronic:2009qf}, including systematic RHIC BES measurements~\cite{STAR:2017sal}.
We parametrize these results over the $\snn$ range considered here as $d\Vfo/dy= a + b\snn + c\snn^2 + d\exp(-\snn/e)$ with $a=908.629$~fm$^3$, $b=12.2579$~fm$^3$/GeV, $c=-0.0299$~fm$^3$/GeV$^2$, $d=69685.5$~fm$^3$, and $e=1.041 \gev$.

We note that in theoretical studies of the QCD phase diagram, the $i$th order susceptibility $\chi_{B,i} \equiv \inparr{^i(P/T^4)}{(\mub/T)^i}|_T$ is defined for (net) baryon number, while the available data is limited to (net) proton moments and cumulants. 
This introduces a mismatch between the experimental observables and susceptibilities defined through pressure derivatives with respect to $\mub$. 
Nevertheless, Ref.~\cite{Hatta:2003wn} argues that the singular parts of the baryon susceptibility~$\chi_B$ and net-proton number fluctuations~$(\delta N_p)^2$ coincide, so that their power laws are the same even if the full susceptibilities differ non-trivially.

Effects due to baryon number conservation can be minimized by avoiding subvolumes large enough to contain more than $\approx 25\%$ of the total proton number~\cite{CAVAGNA20181,boxscaling,Kuznietsov:2022pcn}. 
This condition is met for most points except the widest rapidity bins at $\snn = 2.4$ and $3.0 \gev$, where $W/W_{\rm{max}}$ can reach 0.68. 
These two energies present additional challenges:
1)~Rapidity windows extending from midrapidity to beam rapidity, as for the STAR $\snn = 3.0 \gev$ data, sample regions where $\mufo$ and $\Tfo$ can vary substantially with~$W$.
2) While a strong correlation between coordinate space and momentum space is well-established at $\snn \geq 7.7 \gev$~\cite{Sorensen:2009cz,STAR:2016vqt}, for low energies this mapping becomes more complex so that~$W$ may be a poorer proxy for system size.
3) For the $\snn = 2.4 \gev$ data, there is ambiguity in the extraction of $\Tfo$, $\mufo$, and $\Vfo$~\cite{Motornenko:2021nds} which yields large uncertainties on~$\chi_2$.
We use the HADES experiment values~\cite{HADES:2020wpc} which yield $\Vfo\Tfo^3 = 2.09~{\rm GeV^3\,fm^3}$, however, values in the literature can span $0.46$ to $4.3~{\rm GeV^3\,fm^3}$~\cite{Motornenko:2021nds,Chatterjee:2015fua}. 
Consequently, the $\snn = 2.4$ and $3.0 \gev$ data are more difficult to interpret than those at $\snn \geq 7.7 \gev$.

\section{Non-critical scaling}
\label{sec:non-critical_scaling}

The residual factor~$F$, Eq.~\eqref{eq:non-critical_part_of_the_scaling_function}, allows an analytical determination of the optimal $\mu_c$ for which $F(\mu_B, W) \approx \rm const$, provided that the dependence of $C_1$, $C_2$, and $n_P$ on $W$ and $\mufo$ is known.
We will take $T$ and $\mu_B$ to be given by the chemical freeze-out parameters~$\Tfo$ and $\mufo$.
Since~$\Tfo$ and $\mufo$ are typically extracted only for a single~$W$, we are effectively forced to neglect the $W$-dependence of~$F$.
Then, the condition that~$F$ is flat as a function of~$u$ is equivalent to demanding that it is flat as a function of $\mu_B$,
\begin{align}
\hspace{-2mm}\frac{d F}{d u} = \frac{d\mu_B}{du} \frac{d F}{d\mu_B} \propto  W^{1/\nu} \frac{d F}{d\mu_B} \approx 0 ~~ \Leftrightarrow ~~ \frac{d F}{d\mu_B} \approx 0 ~.
\end{align}
Assuming an ideal hadron-resonance gas description in the Boltzmann approximation, the net proton density is
\begin{align}
n_P %= f_P(T) \Big[e^{\mu_B/T} - e^{-\mu_B/T} \Big] 
= 2f_P(T) \sinh \Big(\frac{\mu_B}{T}\Big) ~,
\end{align}
with $f_P(T) = g \frac{m_P^2 T}{2 \pi^2} K_2\big(\frac{m_P}{T}\big)$ where $m_P$ is the proton mass and $K_2$ is the modified Bessel function of the second kind of order 2.

\begin{figure}[t!] % t = top, b= bottom, h = here, ! = enforce
	\centering
	\includegraphics[width=0.48\textwidth]{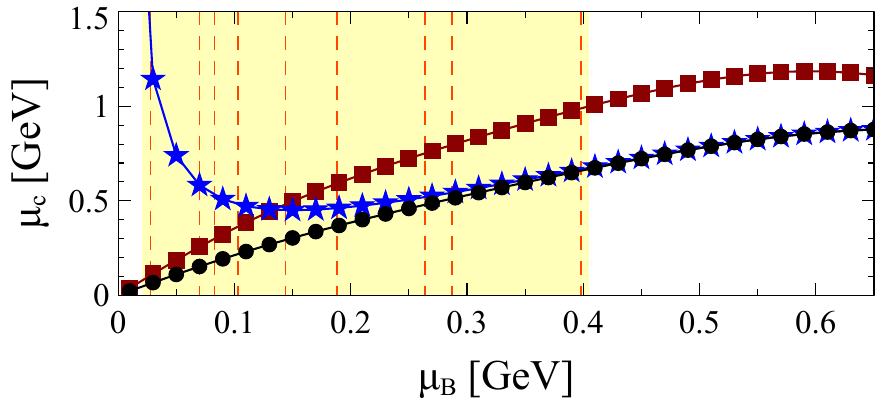} 
 	\caption{Values of $\mu_c$ that make $F$ locally flat for Poisson (black circles) and Skellam (blue stars) distributions, and of~$\chi_3$ for Skellam (maroon squares). 
    Yellow shading marks the BES $\mub$ range, dashed red lines indicate measured $\mufo$ values.
	}
	\label{fig:accidental_scaling_locally_optimal_muc}
\end{figure}

\textbf{Poisson distribution.}
In this case $C_1 = C_2$, giving
\begin{align}
F^{(\rm P) } = \frac{ 2f_P (T)\sinh \big(\frac{\mu_B}{T}\big) }{\big[T( \mu_B)\big]^3}\bigg(1 - \frac{\mu_B}{\mu_c} \bigg)^{\gamma} ~.
\end{align}
The condition $dF/d\mu_B = 0$, equivalently $d(\ln F)/d\mu_B = 0$, yields
\begin{align}
\mu_c = \mu_B +   \frac{ \gamma}{  \frac{x }{  T} \frac{K_1 (x)}{ K_2(x) }\derr{T}{\mu_B}  + \frac{\coth(\frac{\mu_B}{T})}{T} \Big[1 - \frac{\mu_B}{T} \derr{T}{\mu_B}  \Big]} ~,
\label{eq:extracted_critical_mu_c_Poisson}
\end{align}
where $x = m_P/T$.
Here, $\mu_c$ should be understood as a parameter which ensures~$dF/\mu_B = 0 $ for a given~$\mu_B$, \textit{i.e.}, makes~$F$ \textit{locally} flat.
Using the freeze-out parametrization of Ref.~\cite{Andronic:2017pug},
\begin{align}
T (\mu_B) = \frac{0.1584}{1 + \exp\Big[ 2.6 - \frac{\ln \big(\sqrt{s}(\mu_B)\big)}{0.45}  \Big]}  ~,
\end{align}
with $\sqrt{s}(\mu_B) = 4.54/\mu_B - 1/0.288$,
gives
\begin{align}
\frac{dT}{d\mu_B} = -  63.6925 \frac{T^2}{\mu_B^2}  \exp\bigg[ 2.6 - \frac{\ln \big(\sqrt{s}\big)}{0.45}  \bigg]  \frac{1}{\sqrt{s} }~.
\end{align}
This allows one to obtain locally optimal $\mu_c$ which is found to grow monotonically with~$\mu_B$, as shown in Fig.~\ref{fig:accidental_scaling_locally_optimal_muc}.

We determine the \textit{globally optimal} $\mu_c^{(P)}$ by minimizing the reduced standard deviation of~$F$ at the experimentally measured values of~$\mufo$, obtaining~$\mu_c^{(P)} = 0.48\gev$.
Minimizing $\sigma(F)/\langle F\rangle$ over the full BES $\mu_B$ range yields a similar value, $\mu_c = 0.47\gev$.

\textbf{Skellam distribution.}
In this case, we have
\begin{align}
\frac{C_2}{C_1} = \coth \Big( \frac{\mu_B}{T} \Big)~,
\end{align}
so that 
\begin{align}
F^{(\rm Sk) } = \frac{ 2f_P (T)\cosh \big(\frac{\mu_B}{T}\big) }{\big[T( \mu_B)\big]^3}\bigg(1 - \frac{\mu_B}{\mu_c} \bigg)^{\gamma} ~.
\end{align}
Repeating the above calculation then yields
\begin{align}
\mu_c  = \mu_B +   \frac{ \gamma}{\frac{x }{T} \frac{K_1 (x)}{ K_2(x) }  \derr{T}{\mu_B}  + \frac{\tanh (\frac{\mu_B}{T})}{T} \bigg[1 - \frac{\mu_B}{T} \derr{T}{\mu_B}  \bigg]} ~,
\label{eq:extracted_critical_mu_c_Skellam}
\end{align}
which is likewise shown in the upper panel of Fig.~\ref{fig:accidental_scaling_locally_optimal_muc}.
In contrast to Poisson, the locally optimal~$\mu_c$ for Skellam develops a minimum within the BES range, suggesting that Skellam fluctuations may be more susceptible to accidental scaling.
The globally optimal value is obtained as in the Poisson case, yielding $\mu_c^{(Sk)} = 0.55 \gev$ from measured $\mufo$ values and $\mu_c^{(Sk)} = 0.54 \gev$ when minimizing over the full BES range.

We note that the sole difference in~$F$ between the Poisson and Skellam cases is the appearance of $\sinh(\mu_B/T)$ and $\cosh(\mu_B/T)$, respectively, which drives their different scaling behavior.
Near~$\mu_B = 0$, $\sinh (\mu_B/T) \approx \mu_B/T$, so that the Poisson contribution starts from zero and varies strongly with~$\mu_B$.
In contrast, $\cosh (\mu_B/T) \approx  1 + \frac{1}{2}(\mu_B/T)^2$ is much flatter, as seen in the upper panel of Fig.~\ref{fig:accidental_scaling}.
The decreasing factor~$\big( 1 - \mu_B/\mu_c)^{\gamma}$ can therefore compensate for the more gentle variation in $\cosh (\mu_B/T)$ over a broader range in~$\mu_B$, allowing~$F^{(\rm Sk)}$ to remain approximately constant.

For $\chi_3$, the residual factor~$F$, Eq.~\eqref{eq:non-critical_part_of_the_scaling_function}, becomes
\begin{align}
F(W, \mufo) = \frac{C_3(W, \mufo) n_P(W, \mufo)}{ C_1(W, \mufo) \Tfo^{3}} \Big(1 - \frac{\mufo}{\mu_c} \Big)^{\gamma_3}~,
\end{align}
where $\gamma_3$ is the critical exponent associated with~$\chi_3$.
For a Skellam distribution, $C_3/C_1 = 1$, so that we can apply the Poisson result after replacing $\gamma$ with $\gamma_3$. 
The corresponding results are shown in the upper panel of Fig.~\ref{fig:accidental_scaling}.
The globally optimal values are $\mu_c^{(Sk3)} = 0.65 \gev$ for the measured $\mufo$ values and $\mu_c^{(Sk3)} = 0.66 \gev$ over the full BES range.

Higher-order Skellam cumulants follow the same pattern, with even and odd orders behaving like $C_2/C_1$ and $C_3/C_1$, respectively, albeit with different critical exponents.
In particular, we then expect~$\chi_4$ in the Skellam case to be similarly prone to accidental scaling as~$\chi_2$.

\section{Details of codes and data}
\label{sec:details_of_codes_and_data}

Simulations were performed using \texttt{SMASH-3.3}~\cite{SMASH_3_3}.
The code was run in the cascade mode, with no mean field potentials, and with one test particle per physical particle, $N_T = 1$. 
We note that generally, mean fields are needed to quantitatively describe collective observables such as flow, particularly at low beam energies.
Their use, however, requires $N_T > 1$, which is problematic for interpretation of event-by-event fluctuations.
Thus, our calculations should be treated as a qualitative baseline.

At each $\snn$, we generated $3\times 10^5$ events for multiplicity analysis, sampling the impact parameter~$b$ from the interval~$[0, 18.0] \fm$.
We also generated $3 \times 10^6$ events for cumulant analysis, with $b$ sampled from~$ [0, 4.5] \fm$, corresponding roughly to the 0--10\% centrality class.

%Outputs of \texttt{SMASH} simulations were analyzed using an analysis suite for \texttt{SMASH} ROOT output~\cite{SMASH_ROOT_analysis} \as{update the suite, get Zenodo, get a DOI}.
Multiplicity analysis was performed using all charged particles except protons, antiprotons, and deuterons, with a $p_T$-cut of $0.2 \gev < p_T$ and a pseudorapidity cut of $|\eta| < 1$ and $0.05 < \eta < 2$ in the collider and FXT mode, respectively. 
Cumulant analyses were performed for the \mbox{0--5\%} centrality class, with events selected based on multiplicity cuts established in the multiplicity analyses.
Cumulant analyses used the transverse-momentum cut of $0.4 \gev < p_T < 2 \gev$ and applied the centrality bin width correction, using centrality intervals of approximately 1\%.

We analyzed the data using three independent implementations of the FFS analysis. One determined the optimal FSS parameters by minimizing $\chi^2$-square of power-law fits to the scaled~$\chi_2$, second used log-variance-based minimization, and third employed a data collapse measure based on the method of Ref.~\cite{Bhattacharjee:2001dataCollapse}, simplified to linear interpolation.
The results discussed above were insensitive to the method used. 
The second and third implementation are publicly available~\cite{Sorensen_FSS_net_proton_cumulants_2026}.

\end{document}